\documentclass[twocolumn,preprintnumbers,amsmath,amssymb]{revtex4} %
\usepackage{epsfig,graphicx}
\usepackage{dcolumn}
\usepackage{bm}

\begin{document}
\draft

\title{Enhancing Transport Efficiency by Hybrid Routing Strategy}
\author{J.-Q. Dong$^{1}$, Z.-G. Huang$^{1}$\footnote{For correspondence: hangzg@lzu.edu.cn},  Z. Zhou$^{1}$, L. Huang$^{1}$,
Z.-X. Wu$^{1}$, Y. Do$^{2}$, and Y.-H. Wang$^{1}$}
\address{$^{1}$Institute of Computational Physics and Complex Systems, Lanzhou University, Lanzhou Gansu 730000, China}
\address{$^{2}$Department of Mathematics, Kyungpook National University, Daegu, 702-701, Korea}
\date\today

\begin{abstract}
Traffic is essential for many dynamic processes on real networks,
such as internet and urban traffic systems. The transport
efficiency of the traffic system can be improved by taking full
advantage of the resources in the system. In this paper, we
propose a dual-strategy routing model for network traffic system,
to realize the plenary utility of the whole network. The packets
are delivered according to different ``efficient routing
strategies'' [Yan, et al, Phys. Rev. E 73, 046108 (2006)]. We
introduce the accumulate rate of packets, $\eta$ to measure the
performance of traffic system in the congested phase, and propose
the so-called equivalent generation rate of packet to analyze the
jamming processes. From analytical and numerical results, we find
that, for suitable selection of strategies, the dual-strategy
system performs better than the single-strategy system in a broad
region of strategy mixing ratio. The analytical solution to the
jamming processes is verified by estimating the number of jammed
nodes, which coincides well with the result from simulation.
\end{abstract}


\maketitle

\section{Introduction}

Recently, the real transportation or communication systems such as
the computer networks \cite{Albert1999,Pastor2001}, power grid
\cite{WattsPowergrid,power}, airport line \cite{airport}, and so
on, have attracted a lot of attention from scientists due to the
discovery of the topological features of their self-induced
structures. The complex network theory
\cite{Albert2002,Newman2003,Boccaletti}, as well as the tools
inherited from nonequilibrium statistical physics
\cite{Dorogovtsev2008} have been successfully applied to study the
dynamical properties of these real systems.

The common character for these transportation or communication
systems is to perform certain functions by transferring objects
among connected elements, which often take the form of large
sparse network. Free traffic flow on these networks is key to
their normal and efficient functioning. However, they may actually
suffer from the overload or traffic jam, which always disable the
system partially for a period of time, or even be fatal to the
whole system due to the consequential onset of cascades of
overload failures \cite{Strogatz2001,Jacobson1988,Motter2002,
Watts2002PNAS,Moreno2002,Holme2002pre}. Therefore, many recent
studies on the traffic networks have analyzed the critical
properties of the jamming and congestion transitions
\cite{Arenas2001prl,Ohira1998,Sole2001,Valverde2002,Guimera2002prl,Guimera2002pre,Tadi2002,Toroczkai2004,Kujawski2006,Tadi2007,Sreenivasan2007}.
And, the schemes to promote the performance of traffic systems are
chiefly from \emph{two aspects}, designing efficient routing
strategies
\cite{Ramasco2010pre,Wang2006pre,JGG2008pre,Danila2006pre,Goh2001prl,Echenique200405,yan2006,Wu2008jstat,Wang2006pre2,Ling2010pre}
or, optimizing the topology of the underlying network
\cite{Guimera2002prl,Zhang2007pre,Liu2007pre,Zhao2005preLai,Cholvi2005preR}.
The objectives of these schemes are, on one hand, to avoid the
onset of congestion and, on the other hand, to have short delivery
times.

The routing algorithm proposed in recent works are relied on the
structural properties, as well as the global or local information
about the dynamical state of the communication networks
\cite{Ramasco2010pre,Wang2006pre,JGG2008pre,Danila2006pre,Goh2001prl,Echenique200405,yan2006,Wu2008jstat,Wang2006pre2,Ling2010pre}.
For example, the works of biased random walk scheme introduce the
probability to visit node depending on its degree
\cite{Wang2006pre,JGG2008pre}, or the queue length of packets
\cite{Danila2006pre}. The works of shortest-path scheme consider
the paths with minimized distance from any pair of source and
destination \cite{Goh2001prl}. For this scheme, the central nodes
(with highest connectivity) are highly overcongested, inducing the
bottleneck of the communication capacity. The expended version of
the shortest-path scheme with ``effective distance'' involving the
congestion state (queue length of routers) may bypass the
congested nodes locally and thus improve the performance
\cite{Echenique200405}. While, the work of efficient-path schemes
\cite{yan2006} propose the routing table of paths with the minimum
summary of $k^\beta$, with a turnable $\beta$. For the value of
$\beta=-1$ this scheme can effectively redistribute the heavy load
on central nodes to some of the lower-degree nodes, and the system
can reach a more than ten times high capacity of that with
shortest-path scheme. We can see that, for certain amount of
traffic request, the way to promote the performance of the system
is to take full advantage of all kinds of nodes.

These aforesaid researches, have discussed the system with pure
routing strategy. While, how the diversity of routing strategy
performs is really of curious, and the enhancement of transport
capacity by better exertion of all nodes in the system might be
expected. In this paper, we put forward a mechanism that the
communication system possesses of two different routing
strategies. Here we make use of the simple fixed routing scheme,
i.e., the efficient-path schemes proposed in Ref. \cite{yan2006},
and consider the routing strategies to be denoted by different
$\beta$. Then, the transport system with this multi-strategy
protocol will send packets according to different fixed routing
tables of efficient-path schemes. Though the fixed routing
algorithm becomes impractical in huge communication systems, it is
still widely used in medium-sized or small systems
\cite{Tanenbaum1996,Huitema2000}, for its obvious advantages in
economical and technical costs, compared with the dynamical
routing algorithm and information feedback mechanism. In this
case, the diversity of the fixed routing strategy is, of course,
practical if it performs better than pure-strategy system.
Actually, through our study, we see that the multi-strategy system
may perform better than that of the pure strategy system.

\section{TRAFFIC MODEL}

In our traffic model of dual-strategy routing protocol, the
packets with given sources and destinations will be sent according
to two different fixed routing tables of efficient-path schemes
(EPS). For the EPS proposed in Ref.\cite{yan2006}, node $i$ in the
graph are weighted by $w_{i}=k_{i}^{\beta}$. $k_{i}$ is the degree
of node $i$, and $\beta$ can be considered as the label of
``routing strategy''. A packet with source $j_{1}$ and destination
$j_{2}$ will choose a minimum sum of weight, $\sum_{i\in
\sigma_{j_{1}j_{2}}}{k_{i}^{\beta}}$, route in the graph.
$\sigma_{j_{1}j_{2}}$ is the path from $j_{1}$ to $j_{2}$.
Adjusted by the parameter $\beta$, the single-strategy system will
partial to certain kind of nodes in routing, and may also leave
some space to improve the performance further. In our
dual-strategy model with two strategies $\beta_{1}$, and
$\beta_{2}$, packets are assigned to the two corresponding routing
tables, with probability $1-p$ and $p$, respectively. Here we name
$p$ as the \emph{mixing rate}. Here, for $p=0$ (or $1$), the
system returns to the single-strategy system with $\beta=\beta_1$
(or $\beta_2$).

Similar to the former work, at each time step, $R$ packets enter
the system with randomly chosen sources and destinations. The
delivery capacity of each node is $C$, and we set $C=1.0$ for
simplicity. The maximal queue length of each node is assumed to be
unlimited, and the first-in-fist-out discipline is applied at each
queue. Once a packet come to its destination, it is removed from
the system.

In the previous study, the phase transition of traffic flow is
described by the the order parameter \cite{Arenas2001prl},
\begin{equation}
H(R) = \lim_{t \rightarrow \infty}\frac{C}{R}\frac{\langle \Delta
W\rangle}{\Delta t} \label{eq0}
\end{equation}
where $\Delta W=W(t+\Delta t)-W(t)$, with $\langle\cdot\rangle$
indicating average over time windows of width $\Delta t$, and
$W(t)$ is the total number of packets in the network at time $t$.
The critical value $R_c$ (the packet generation rate) where a
phase transition takes place from free flow to congested traffic,
can reflect the maximum capability of a system.

The behavior of the critical point $R_c$ on different networks can
be simply explained by their different betweenness centralities
(BC) distributions \cite{Newman2001pre,Newman2004pre,Goh2001prl}.
The BC of a node $i$ for the single-strategy EPS system
\cite{yan2006} is defined as,
\begin{equation}
g_i(\beta)=\sum_{j_{1}\neq j_{2}} \frac{\sigma_{j_{1}
j_{2}}(\beta,i)}{\sigma_{j_{1} j_{2}}(\beta)}, \label{eq3}
\end{equation}
where $\sigma_{j_{1} j_{2}}(\beta)$ is the number of routes going
from $j_{1}$ to $j_{2}$, according to the EPS routing table with
$\beta$; While, $\sigma_{j_{1} j_{2}}(\beta,i)$ is the number of
those also passing through $i$. The critical value $R_c$ can be
estimated by the maximal BC as,
\begin{equation}
R_{c}=\frac{C\cdot N\cdot (N-1)}{Max[g_i(\beta)]}. \label{eqRc1}
\end{equation}
where $Max[g_i(\beta)]$ is the maximal $BC$ of the system with
strategy $\beta$.

For the dual-strategy system with strategies $\beta_{1}$,
$\beta_{2}$, and probability $p$, the efficient BC of one given
node $i$ is,
\begin{equation}
G_i(\beta_{1},\beta_{2},p)=(1-p)\cdot g_i(\beta_{1})+p\cdot
g_i(\beta_{2}) \label{eq5}
\end{equation}
Then, we have the load of node $i$, assigned from the whole
transport requirement of the system as,
\begin{equation}
L_i=\frac{G_i(\beta_1,\beta_2,p)\cdot R}{N\cdot (N-1)}\label{eqLi}
\end{equation}
The load of node increases as the $R$ is increased. Therefore, the
critical value $R_c$ can be estimated as,
\begin{equation}
R_{c}=\frac{C\cdot N\cdot (N-1)}{Max[G_i(\beta_{1},\beta_{2},p)]},
\label{eqRc2}
\end{equation}
here, $Max[G_i(\beta_{1},\beta_{2},p)]$ is the maximal efficient
BC of the dual-strategy system.

\begin{figure}
\small \centering
\includegraphics[width=8cm]{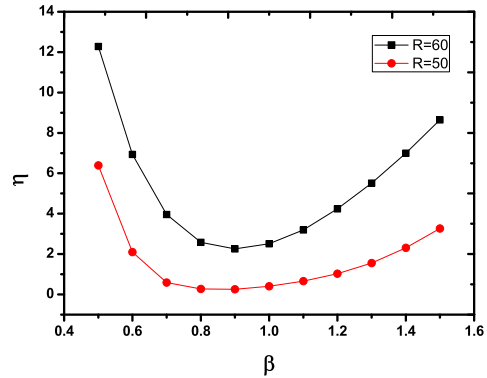}
\caption{The accumulate rate $\eta$ as a function of $\beta$ in
the single-strategy system, for the systems of $R=50$ and $60$.
The results are averaged over $10$ realizations for $20$ networks,
with size $N=1225$.}\label{fig1}
\end{figure}

\section{SIMULATION RESULT AND ANALYSIS}

The communication networks typically show a scale-free (SF)
distribution for the number of links departing from and arriving
to a system element. In this paper, we choose  Barab\'{a}si-Albert
(BA) network as the communicating network \cite{ba}. For this
network model, starting from $m_{0}=3$ fully connected nodes, new
node with $m=2$ is added in the existing network in turn, until
the network size $N=1225$. The network average degree $\langle
k\rangle=4$.

For the single-strategy system, the phase transition from free
flow to congested traffic has been discussed \cite{yan2006}. When
the value of $R$ increases over $R_{c}$, the number of accumulated
packets get to increase with time (i.e., a phase transition takes
place from free flow to congested traffic). Similarly, for the
multi-strategy system, the phase transition also takes place. The
effect of different strategies, in free flow phase,  is merely
inducing the difference of packet deliver time. While in the
congested phase, much more diversified phenomenon appears. We
mainly focus on the congested phase as follows.

Firstly, let us revisit the behavior of the single-strategy system
in the congested phase. According to the work of Yan
\cite{yan2006}, the largest $R_{c}$ (around $43$), i.e. the best
performance of the system, is achieved with strategy $\beta=1.0$
on BA network of $N=1225$ and $\langle k\rangle=4$. From
systematic simulation of various $\beta$ systems in congested
phase, we notice that the number of accumulated packets increases
linearly with $t$. Namely, the accumulate rate $\eta$ is a
constant (with small fluctuation). In Fig. \ref{fig1}, we shows
$\eta$ as a function of strategy $\beta$, with $R$ in the region
of congested phase ($R=50$ and $60$, larger than $R_{c}$). It is
necessary to emphasis that, although the so-called congestion
occurs, there still are, on average, $R-\eta$ packets successfully
delivered to their destinations per unit time. This number is
actually much larger than $R-R_c$. That is to say, while some
nodes are jammed as $R>R_c$, a noticeable part of transport
function still holds in the system. This actually is realized from
two aspects, $(1)$ the ``free flow'' still takes place on the
paths which are not entangled with the jammed nodes, and, $(2)$
the packets through the jammed nodes are not stopped but just
delayed.

We may say that, the parameter $R_{c}$ merely distinguishes the
so-called free and congested phases, which actually indicates the
free or jammed state of the most ``fragile'' node [see Eq.
(\ref{eqRc1})]. $R_{c}$ can not reflect the extent of congestion,
and the impact of the jammed nodes to the performance of the
system. However, the accumulate rate, defined as,
\begin{equation}
\eta= \lim_{t \rightarrow \infty}\frac{\Delta W}{\Delta t},
\label{eq7}
\end{equation}
is a good parameter to measure the performance of the system in
the congested phase. The smaller $\eta$ denotes better performance
of the system. $\eta$ is the sum of individuals' $\eta_{i}'$ over
the whole system as, $\eta=\sum_i{\eta_{i}'H(\eta_{i}')}$. Here,
$H(\cdot)$ is the Heaviside function, and $\eta_{i}'$ is the
individual accumulate rate of node $i$, namely, the increase rate
of the queue length of packets at node $i$ per time step.

From Eq. (\ref{eqLi}), we can get the analytically expression of
$\eta'_i$ as,
\begin{equation}
\eta'_i\equiv L_i - C=\frac{G_i(\beta1\beta2,p)\cdot R}{N\cdot
(N-1)}-C \label{eq8}
\end{equation}
with $L_i$ the load of node $i$ assigned from the whole transport
requirement. We may notice that as $R$ is increased, $L_i$ may
increases over the capability $C$ and thus $\eta'_i$ increases
from negative to positive.

In Fig. \ref{fig1}, the non-monotonic behavior of $\eta$ implies
that the medium $\beta$ system performs better, similar to the
results in Ref. \cite{yan2006} from the relationship between
$R_{c}$ and $\beta$.

\begin{figure}
\small \centering
\includegraphics[width=8cm]{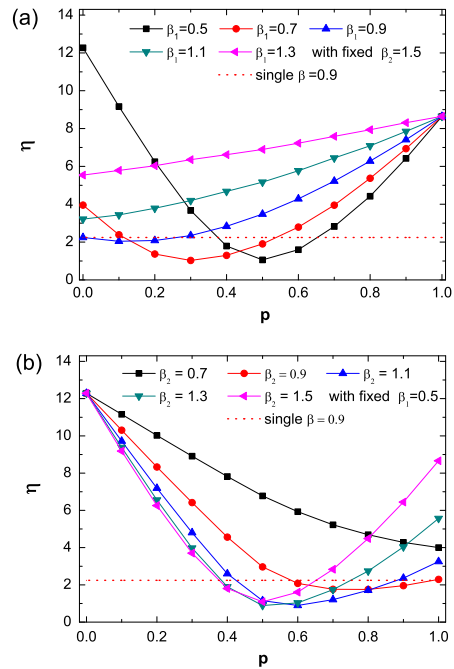}
\caption{(Color online.) The accumulate rate $\eta$ for the
dual-strategy system as a function of \textbf{mixing ratio} $p$ of
the two strategies $\beta_{1}$ and $\beta_{2}$. Here, in (a),
$\beta_{2}$ is fixed to be $1.5$, and in (b), $\beta_{1}$ is fixed
to be $0.5$. The $\eta$ of the single-strategy system with optimal
$\beta=0.9$ (the red dot line) is also plotted for comparison. The
results shown are averaged over $10$ realizations for $20$
networks, with size $N=1225$, and $R=60$.}\label{fig2}
\end{figure}

Then, we will analysis the behavior of the dual-strategy system
with $\beta_{1}$ and $\beta_{2}$ in the congested phase. The
packets are assigned to the two strategies with probability $1-p$
and $p$, respectively. Figure \ref{fig2} plots $\eta$ of the
system as a function of $p$. Here, for $p=0$ (or $1$), $\eta$
returns to that of the single-strategy system with $\beta=\beta_1$
(or $\beta_2$). We can see that, the mix of different strategies
is nontrivial and of interest. Take the system with
$\beta_{1}=0.5$ and $\beta_{2}=1.5$ in Fig. \ref{fig2}(a) as an
example, for certain medium value of $p$, it performs even better
than the optimal state the single-strategy system achieves with
$\beta=0.9$ (which is also plotted by the red dot line in Fig.
\ref{fig2}). Furthermore, as has been shown in Fig.  \ref{fig2},
it is also noteworthy that, when $\beta_{1}$ and $\beta_{2}$ are
chosen from each side of $0.9$, there always exists an optimal
configuration $p$, which performs better both than the
single-strategy systems of $\beta_{1}$ and $\beta_{2}$.

This can be understand as follows. To design routing strategy for
the network transportation, there are two factors that should be
considered. (1) To bypass the hub nodes which are obviously of
heavy burden and prone to jamming. (2) To choose shorter path to
reducing deliver time, which is conducive to reduce the occupation
(life time) of packets to the resources and thus avoid jam. The
system deliver efficiency can be improved from the tradeoff of
these two factors. However, they are inconsistent in the
communicating network with heterogeneous topology. Take the
single-strategy system in congested phase as an example (see Fig.
\ref{fig1} the curve with $R=60$), as $\beta$ is increased from
$0$, the traffic through the hub nodes are bypassed to the other
smaller degree nodes, while the lengthes of the pathes adopted are
prolonged, which increases the probability of jamming for the
other nodes. The system with $\beta_0$ around $0.9$, to certain
extent, is compatible of these two factors, and thus achieves the
optimal performance. As $\beta$ is increased further, the utility
of the hubs is not sufficient, while the left parts of the system
are overworked. Actually, To take a full advantage of each node in
the system will return better performance. Therefore, for the
dual-strategy system, the strategy inclined to the hubs
($\beta<\beta_0$) and that inclined to the small nodes
($\beta>\beta_0$) may complement each other and perform better
than the single strategy one. Thus non-monotonous $\eta$ can be
observed when the $\beta$ from both side of $\beta_0$ are mixed.

The effect of multiple strategies in the congested phase can also
be understand analytically from the so-called \emph{equivalent
generation rate}. In this routing strategy, packets at the head of
the queue on node $i$ will be delivered to the next node $j$
according to the routing table, no matter node $j$ is idle or
jammed. Current server also has this properties. In this case,
congestion in the system will not spread out. Furthermore,
counterintuitive, congestion will make the system more ``empty''.
In each time step, $\eta$ more packets will queue at the jammed
nodes, and as a consequence, the load of the other nodes will be
lighten, as if the generation rate for the \emph{subsystem} of
these nodes is reduced to a smaller one $R^{*}$, which we name as
the \emph{equivalent generation rate}. Here, we have
\begin{equation}
R^{*}=R-\eta. \label{eqReat}
\end{equation}
Different from the case that the servers abandon packets when the
queue length is over a threshold, in our model, the queuing
packets are not abandoned, and will finally be send to their
destination.

We sort nodes by the values of their individual accumulate rates
in descending order, as $\eta'_1>\eta'_2>...>\eta'_N$. From Eq.
(\ref{eq8}), we know that, when $R$ is increased from $0$, all
these $\eta'_i$ increases from $-C$. As soon as the maximum one,
$\eta'_1$, increases from negative to positive, the system
transform from free phase to congested phase. Suppose that
$\eta'_2<0$, there are $\eta'_1$ packets detained at the $1$st
node per time step. Then, the equivalent generation rate for the
subsystem (exclude the $1$st node) is $R^{*}=R-\eta'_1$. As $R$ is
increased further, the left nodes will be jammed one after another
(i.e., have positive $\eta'_i$). Accordingly, we may propose the
\emph{theory} to predict the number of jammed nodes, and the
accumulate rate of the system $\eta$ from two perspective.

\begin{figure}
\small \centering
\includegraphics[width=8cm]{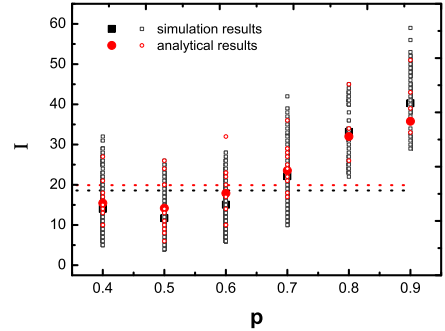}
\caption{The number of jammed nodes from analytical and simulation
results, for the dual-strategy system with
$\beta_{1}=1.5,\beta_{2}=0.5$. The sample data of analytical
results (red open circle) are from $10$ different networks, and
that of simulation results (black open square) are from $50$
realizations of traffic on these $10$ networks. The average number
of analytical and simulation results (red solid circle and black
solid square) are averaged over the corresponding sample data. The
system is of size $N=1225$, and $R=60$. The analytical and
simulation results (the red and black dot lines) from
single-strategy system with $\beta=0.9$ are also plotted for
comparison.}\label{fig3}
\end{figure}

On one hand, from Eqs. (\ref{eq8}) and (\ref{eqReat}), we get,
\begin{equation}
R^{*}=R-\sum_{i=1}^{I}[\frac{G_i(\beta_{1},\beta_{2},p)\cdot
R^{*}}{N\cdot(N-1)}-C]\label{eq:R1}
\end{equation}
with the following constraint applies:
\begin{eqnarray}
&&L_{I}=\frac{G_{I}(\beta_{1},\beta_{2},p)\cdot R^{*}}{N\cdot(N-1)}>C,\label{eq:LI}\\
&&L_{I+1}=\frac{G_{I+1}(\beta_{1},\beta_{2},p)\cdot
R^{*}}{N\cdot(N-1)}<C.\label{eq:LI+1}
\end{eqnarray}
By solving this problem, we can get the number of jammed nodes
$I$, and $\eta$, for given values of $R$, $\beta_{1}$, $\beta_{2}$
and $p$.

On the other hand, we focus on the detailed process of
successional jamming which gradually modifies the equivalent
generation rate $R^*$, as well as the load $L_i$ of the left
nodes. The iterative procedure of $R^*$ can be written as,
\begin{eqnarray}
&&R^{*}_1=R-\frac{G_1(\beta_{1},\beta_{2},p)\cdot R}{N\cdot(N-1)}+C\\
\nonumber && R^{*}_2=R^{*}_1-\frac{G_2(\beta_{1},\beta_{2},p)\cdot
R^{*}_1}{N\cdot(N-1)}+C\\ \nonumber
 &&... ... ...
\end{eqnarray}
The iterative formula is,
\begin{equation}
R^{*}_i=R^{*}_{i-1}-\frac{G_{i}(\beta_{1},\beta_{2},p)\cdot
R^{*}_{i-1}}{N\cdot(N-1)}+C, (i=1,2,3,...) \label{eq:iterative}
\end{equation}
$R^{*}_i$ and $L'_i$ decrease as the nodes of large load is jammed
one after another, until
\begin{eqnarray}
&&L_{I}'=\frac{G_{I}(\beta_{1},\beta_{2},p)\cdot R^{*}_{I-1}}{N\cdot(N-1)}>C,\label{eq:LIb}\\
&&L_{I+1}'=\frac{G_{I+1}(\beta_{1},\beta_{2},p)\cdot
R^{*}_{I}}{N\cdot(N-1)}<C,\label{eq:LI+1b}
\end{eqnarray}
Different from Eqs. (\ref{eq:R1}) to (\ref{eq:LI+1}), Eqs.
(\ref{eq:iterative}) to (\ref{eq:LI+1b}) depicts that the jamming
of the first $I$ nodes steps down $R^{*}$ gradually until the
value $R^{*}_{I}$, where the $(I+1)^{th}$ node, as well as all its
following nodes, is capable of treating with its load. Here, from
the perspective of successional jamming process described by Eq.
(\ref{eq:iterative}), one can also get the number of jammed nodes
$I$, and $\eta$, analytically.

In Fig. \ref{fig3}, we plot the analytical and simulation results
of the number of jammed nodes $I$ in the dual-strategy system with
$\beta_{1}=1.5$ and $\beta_{2}=0.5$. It can be seen that, the
average number of jammed nodes from analysis (red solid circle)
coincides well with that from simulation (black solid square).
Interestingly, the value of $I$ also behaves non-monotonically and
achieve the minimum around $p=0.5$, which is similar to the
accumulate rate $\eta$ of the same system shown in Fig.
\ref{fig2}. Additionally, the analytical results from Eq.
(\ref{eq:R1}) and Eq. (\ref{eq:iterative}) are very close to each
other, thus in Fig. \ref{fig3} we merely plot the results from Eq.
(\ref{eq:iterative}).

Here, we can also understand the non-monotonic behavior of $I$
from the following perspective. The packet generation rate $R$ can
be divided into two parts, the packets using routing table of
$\beta_1$ is $R^{\beta_1}=(1-p)R$, and that of $\beta_2$ is
$R^{\beta_2}=pR$. From Eq. (\ref{eqLi}), we can get the
corresponding loads of node $i$ from these two parts of packets,
denoted by $L_{i}^{\beta_1}$ and $L_{i}^{\beta_2}$ (with
$L_{i}=L_{i}^{\beta_1}+L_{i}^{\beta_2}$). For the case that the
mixing rate $p=0$, we have $R^{\beta_1}=R$, and the jamming of
nodes are all ascribed to the queue of $\beta_1$ packets. As $p$
is increased from $0$, the $R^{\beta_1}$, as well as the
$L_{i}^{\beta_1}$ decreases, while that of $\beta_2$ increases. If
the $\beta_2$ packets prefer to use those \emph{complementary
nodes} instead of the nodes already jammed by $\beta_2$ packets,
the number of jammed nodes $I$ will decreases with $p$. However,
as $p$ is large enough, the increase of load $L_{i}^{\beta_2}$
from $\beta_2$ packets induces new jamming of nodes. Therefore, we
can see the non-monotonic behavior of the number of jammed node,
when the dual-strategy system is composed of the two strategies
from either side of $\beta_0$.

\section{CONCLUSION}

In summary, we propose a hybrid routing strategy for the networked
traffic system, which is proved to be a doable and effective way
to enhance transport efficiency. Compared with the efficient
routing strategy \cite{yan2006}, the hybrid routing strategy can
make better use of the resources in the traffic system, while
there appears no increase in its algorithmic complexity. The
performance of the dual-strategy system can be optimized by
modulating the mixing rate of the packets, in case that the two
strategies share fewer key nodes. Here, we introduce the
accumulate rate $\eta$ to denote the performance of the
communication system in congestion phase, which shows richer
phenomena than the critical generation rate $R_c$. Furthermore, we
get analytical descriptions to the jamming processes by the
accumulate rate $\eta$ and the equivalent generation rate $R^{*}$.
The number of jammed nodes estimated from analytical formula
coincides well with that from simulation.

While our model is based on computer networks, we expect it to be
relevant to other practical transport processes in general.
Actually, in real system, the hybrid routing is worthy of
considering, for the reason that the sources and characters of
massages delivering or spreading in complex systems are
diversified, which induces the hybrid of various transportation
modes. In view of the common features for the networked traffic
and spreading, our work may shed some light on the research of
packet delivery in technical networks, as well as the rumor and
opinion dynamics in social networks.\\

We gratefully acknowledge T. Zhou and X. Li for helpful
discussions.

\bigskip
\end{document}